\documentclass[aps,pre,twocolumn]{revtex4-1}

\usepackage{graphicx}
\usepackage{dcolumn}
\usepackage{bm}
\usepackage{amsmath}
\usepackage{cases}

\begin{document}

\title{Superlinear and sublinear urban scaling in geographical network model of the city}

\author{K.~Yakubo}
\email{yakubo@eng.hokudai.ac.jp}
\affiliation{Department of Applied Physics, Hokkaido University, Sapporo 060-8628, Japan}
\author{Y.~Saijo}
\affiliation{Department of Applied Physics, Hokkaido University, Sapporo 060-8628, Japan}
\author{D.~Koro\v sak}
\email{dean.korosak@um.si}
\affiliation{University of Maribor, Slom\v skov trg 15,
Maribor SI-2000, Slovenia}

\date{\today}

\begin{abstract}

Using a geographical scale-free network to describe relations
between people in a city, we explain both superlinear and
sublinear allometric scaling of urban indicators that quantify
activities or performances of the city. The urban indicator
$Y(N)$ of a city with the population size $N$ is analytically
calculated by summing up all individual activities produced by
person-to-person relationships. Our results show that the urban
indicator scales superlinearly with the population, namely,
$Y(N)\propto N^{\beta}$ with $\beta>1$ if $Y(N)$ represents a
creative productivity and the indicator scales sublinearly
($\beta<1$) if $Y(N)$ is related to the degree of
infrastructure development. These coincide with allometric
scaling observed in real-world urban indicators. We also show
how the scaling exponent $\beta$ depends on the strength of the
geographical constraint in the network formation.

\begin{description}
\item[PACS numbers] 89.75.Da, 89.65.Lm, 89.65.-s, 89.75.Hc
\end{description}
\end{abstract}

\maketitle

\section{INTRODUCTION}

Cities are often compared to living organisms with a
hierarchical organization consisting of cells, tissues and
organs. Likewise, people in a city form groups, groups form
organizations serving certain functions, and interdependent
complex relationships between functional organizations sustain
the whole urban activities. Such similarities are not only
found in the correspondence between constituent elements of
cities and living organisms but also in allometric scaling. As
is a metabolic rate of a complex organism proportional to the
$3/4$ power of body mass \cite{Rubner83, Kleibe32}, various
quantities related to activities or performances of a city
depend on the scale of the city in a power-law manner
\cite{Nordbeck71,Batty94,Batty08,Batty08b,Batty12,Lo02,Pumain04,
Samaniego08,Chen09,Chen10}. In particular, extensive work
\cite{Glaeser99,Kuhnert06, Bettencourt07, Bettencourt07b,
Bettencourt08,Bettencourt10, Gomez-Lievano12, Alves13, Lobo13,
Bettencourt13a, Changizi10, Arbesman11, Wu11, Alves13,
Alves13b, Fragkias13, Pan13, Lamsal13} has revealed that an
urban indicator $Y$ quantifying city activity scales, on
average, with the population size $N$ as a power-law:
\begin{equation}
Y(N)\propto N^{\beta},
\label{urban_scaling}
\end{equation}
where $\beta$ is a scaling exponent. Bettencourt \textit{et
al.} \cite{Bettencourt07,Bettencourt07b, Bettencourt08} have
found that an urban indicator representing a creative
productivity, such as: the number of new patents, the gross
domestic product (GDP), the number of crimes, etc., obeys a
superlinear scaling law ($\beta>1$) while an indicator related
to the degree of infrastructure development, such as the total
length of electrical cables, the number of gas stations, the
total road surface, etc., scales sublinearly with the
population size  ($\beta<1$). Due to the nonlinear scaling
Eq.~(\ref{urban_scaling}), a meaningful comparison between
characteristics of individual cities requires evaluations of
deviations from this average scaling behavior, instead of
considering per capita quantity $Y(N)/N$ \cite{Bettencourt10,
Gomez-Lievano12, Alves13, Lobo13}.

It is crucial to understand the reason why urban indicators
representing creative productivities scale superlinearly and
those corresponding to material infrastructures scale
sublinearly. Arbesman, Kleinberg, and Strogatz \cite{Arbesman09}
have proposed a network model (AKS model) to explain
superlinear scaling found in creative productivities. They
introduced hierarchical social distances between nodes
representing people in a city. A network is formed by
connecting nodes with the edge probability decaying
exponentially with the social distance. Assuming that the
individual productivity yielded by an edge increases
exponentially with the social distance, the AKS model gives
superlinear scaling of creative productivity $Y(N)$ if the
total contribution from connected node pairs separated by the
social distance $d$ is an increasing function of $d$, giving
linear scaling otherwise.

In order to explain both superlinear and
sublinear scaling of urban indicators, Bettencourt \cite{Bettencourt13b}
has worked with four simple assumptions: (1) Citizens explore the
city fully to benefit from it and the city develops in a way to make this
possible. (2) The infrastructure network volume $A_{n}$ grows in a
decentralized way in order to connect each addition of a new inhabitant,
namely, $A_{n}\propto Nr$, where $N$ is the number of people in the
city and $r$ is the average distance between individuals. (3) The
product of average social output and the volume spanned by individual's
movement is constant of city size $N$, which means that human effort
is bounded. (4) The urban indicator $Y(N)$ related to a creative
productivity is proportional to the number of local social interactions.
According to the Bettencourt's model, the scaling exponent is given by
$\beta=1+\delta$ for the superlinearly scaled creative productivity
and $\beta=1-\delta$ for the sublinearly scaled infrastructure volume,
where $\delta$ is a positive exponent that depends on the fractal
dimension of human travel paths.

Despite these two pioneering and suggestive theories, the mechanism
of urban scaling has not yet been completely understood. Although
the AKS model \cite{Arbesman09} gives a possible explanation
of superlinear (or linear) scaling of creative productivities and
describes how the social structure (i.e., human relations) affects
the scaling exponent $\beta$, sublinear scaling for urban indicators
reflecting infrastructures has not been argued. On the other hand,
the Bettencourt's model \cite{Bettencourt13b} demonstrates both superlinear
and sublinear scaling for a creative productivity and an infrastructure
volume, respectively. However, it is not clear how
the social structure influences urban scaling, because this theory
is based on a continuum model. Furthermore, the scaling exponent
$\beta$ always appears symmetrically as $\beta=1\pm \delta$ for
superlinear and sublinear scaling, hence a variety of real-world
nonlinear urban scaling cannot be described by this model. It is
therefore important to explain consistently both superlinear and sublinear
scaling in the context of the relation between the scaling behavior
and the social structure in the city.

In this paper, we propose a model to account for urban scaling
by representing human relations in a city by a geographical
network in which nodes close to each other are more likely to
be connected. It is assumed that an urban indicator $Y(N)$ is
given by the sum of the activities produced by individual
connected node-pairs and that the individual activity $y_{ij}$
depends on the Euclidean distance $l_{ij}$ between connected
nodes $i$ and $j$. We show that the urban indicator scales
superlinearly or linearly with the population size $N$ when the
activity $y_{ij}$ represents a creative productivity that is an
increasing function of the Euclidean distance $l_{ij}$ and
scales sublinearly or linearly if $y_{ij}$ decreases with
$l_{ij}$ as the strength of the demand for infrastructure does.
This result is consistent with observed urban scaling
phenomena. We also predict that urban indicators representing
either creative productivities or infrastructures are
proportional to the population size (i.e., linear scaling) if
the geographical constraint in the network formation is strong
enough.

The paper is organized as follows. After presenting our model in
Sec.~II, the urban scaling exponent $\beta$ is analytically
calculated in Sec.~III. Numerical confirmations for analytical
results are given in Sec.~IV. We also show here how the
exponent $\beta$ depends on parameters characterizing our model.
Finally, we conclude our work in Sec.~V.

\section{THE MODEL}

\subsection{Geographical network model}

It has been demonstrated that urban structure possesses a self-similar
property, i.e., the fractal nature of population density in a city
\cite{Batty94,Smeed63,Frankhauser98,Yook02,Chen10b}. In our model, $N$
nodes representing people in a city are located homogeneously in a
fractal space $S_{D}$ with the fractal dimension $D$. The Euclidean
distance is defined for any pair of nodes. The fractal space $S_{D}$ with
the linear size $L$ is assumed to be large enough and isotropic from
any point in $S_{D}$. Thus, the number of nodes or the population
size of the city is presented by
\begin{equation}
N= \rho L^{D} ,
\label{node_num_frac}
\end{equation}
where $\rho$ is a coefficient. Each node (person) has its own
ability or charm to attract others. In order to quantify such
personal attractiveness, a real continuous quantity $x$
(referred as `attractiveness' hereafter) is randomly assigned
for each node according to the power-law probability
distribution function $s(x)$ expressed by
\begin{equation}
s(x)=s_{0}x^{-\alpha} , \qquad (x\ge x_{\text{min}})\ ,
\label{sx}
\end{equation}
where $\alpha>1$, $x_{\text{min}}>0$, and the normalization constant
$s_{0}$ is given by
\begin{equation}
s_{0}=(\alpha-1)x_{\text{min}}^{\alpha-1}\ .
\label{s0}
\end{equation}
Since it is natural to consider that two nodes spatially close to each
other and having large attractiveness values are more likely to be connected,
two nodes $i$ and $j$ are connected if the following condition is satisfied,
\begin{equation}
\frac{x_{i}x_{j}}{l_{ij}^{m}}>\Theta \ ,
\label{condition1}
\end{equation}
where $l_{ij}$ denotes the Euclidean distance between the nodes
$i$ and $j$, $m(\ge 0)$ is a parameter controlling the strength
of the geographical constraint in the network formation, $x_{i}$
is the attractiveness of the node $i$, and $\Theta$ is a threshold
value.

Statistical properties of networks formed by the above procedures
have been studied previously \cite{Masuda05,Yakubo11}. We briefly
summarize the results of these works here. First, the network
exhibits the scale-free property, that is, the distribution $P(k)$
of the degree $k$ follows a power law,
\begin{equation}
P(k)\propto k^{-\gamma}\ ,
\label{degree_dist}
\end{equation}
for large $k$ \cite{Masuda05,Yakubo11}. The exponent $\gamma$ is related to
the model parameters $D$, $\alpha$, and $m$ through \cite{Yakubo11}
\begin{equation}
\gamma=
\begin{cases}
2                       & \text{if }D\ge d_{\text{c}}\ , \\[5pt]
\displaystyle
1+\frac{d_{\text{c}}}{D} & \text{if }D<d_{\text{c}}\ ,
\end{cases}
\label{gamma}
\end{equation}
where
\begin{equation}
d_{\text{c}}=m(\alpha-1)\ .
\label{dc}
\end{equation}
This result shows that the degree distribution becomes more homogeneous
when the geographical constraint is enhanced by increasing $m$. This is
because the network formed by a large $m$ value has a lattice-like
structure.

Second, the probability distribution function $R(l)$ of the edge length
$l$ is proportional to the average number of edges, $k(l)dl$, of the
length in the range of $[l,l+dl]$ from a given node. These are given by
\cite{Yakubo11}
\begin{equation}
R(l)\propto k(l)\propto
\begin{cases}
l^{D-1} & \text{if }l\le \xi \ ,\\[2mm]
\displaystyle
l^{D-1}\left(\frac{l}{\xi}\right)^{-d_{\text{c}}} & \text{if }l>\xi\ ,
\end{cases}
\label{kl1}
\end{equation}
where
\begin{equation}
\xi= \left(\frac{x_{\text{min}}^{2}}{\Theta}\right)^{1/m}\ .
\label{xi}
\end{equation}
The quantity $\xi$ is the distance below which any two nodes are connected
regardless of the attractiveness $x$. Here, we neglected a logarithmic
correction term. The probability of two nodes separated by the Euclidean
distance $l$ to be connected by an edge is directly obtained from Eq.~(\ref{kl1}).
This probability $g(l)$ is presented by the ratio of $k(l)dl$ to the
number of nodes $n(l)dl$ located at a distance within the range of
$[l,l+dl]$ from a given node. Since $n(l)\propto l^{D-1}$, the relation
$g(l)=k(l)/n(l)$ immediately leads
\begin{equation}
g(l)=
\begin{cases}
1 & \text{if }l\le \xi \ , \\[2mm]
\displaystyle
\left(\frac{l}{\xi}\right)^{-d_{\text{c}}} & \text{if }l>\xi\ .
\end{cases}
\label{gl1}
\end{equation}
The power-law decay of $g(l)$ for $l> \xi$ is consistent with the
fact that the probability of two persons separated by $l$ to be
socially connected decreases with $l$ in a power-law manner
\cite{Liben-Nowell05,Adamic05,Goldenberg05,Lambiotte08}. The
relation $g(l)=1$ for $l\le \xi$ is obvious from the meaning
of the distance $\xi$. 

Third, the average degree $\langle k\rangle$ of the network can
be controlled by tuning the threshold $\Theta$. Although the
$\Theta$ dependence of $\langle k\rangle$ has been already
studied \cite{Yakubo11}, here we clarify not only the $\Theta$
dependence but the $N$ dependence of $\langle k\rangle$. The
average degree is obviously given by
\begin{equation}
\langle k\rangle = \int_{0}^{L} k(l)\,dl\ ,
\label{kav1}
\end{equation}
where the linear size $L$ of the city is related to the population
size $N$ through Eq.~(\ref{node_num_frac}). Substituting Eq.~(\ref{kl1})
into Eq.~(\ref{kav1}), $\langle k\rangle$ can be calculated as
\begin{eqnarray}
\langle k\rangle &=& c_{1}\int_{0}^{\xi} l^{D-1}\,dl +c_{2}\int_{\xi}^{L} \left(\frac{l}{\xi}\right)^{-d_{\text{c}}} l^{D-1}\,dl \nonumber \\[5pt]
&=& \left(\frac{c_{1}}{D}-\frac{c_{2}}{D-d_{\text{c}}}\right)\xi^{D}+\frac{c_{2}}{D-d_{\text{c}}}\xi^{d_{\text{c}}}L^{D-d_{\text{c}}} \ ,
\label{kav2}
\end{eqnarray}
where $c_{1}$ and $c_{2}$ are irrelevant numerical coefficients.
Here, we define a new relation symbol ``$\propto:$" to represent the
relation $A=cx+c'y$ by $A\propto: x+y$ if $c$ and $c'$ are nonzero
constants independent of $x$ and $y$. Using this notation,
Eq.~(\ref{kav2}) can be written as $\langle k\rangle\propto: \xi^{D}+\xi^{d_{\text{c}}}
L^{D-d_{\text{c}}}$. Thus, the relation $L\propto N^{1/D}$ from
Eq.~(\ref{node_num_frac}) and Eq.~(\ref{xi}) lead
\begin{equation}
\langle k\rangle \propto: \Theta^{-D/m}+\Theta^{-d_{\text{c}}/m}N^{1-d_{\text{c}}/D}\ .
\label{kav3}
\end{equation}
Therefore, we obtain
\begin{equation}
\langle k\rangle \propto
\begin{cases}
\Theta^{-D/m} & \text{if } D\le d_{\text{c}} \ ,\\
\Theta^{-d_{\text{c}}/m}N^{1-d_{\text{c}}/D} & \text{if } D>d_{\text{c}}\ ,
\end{cases}
\label{kav4}
\end{equation}
for a large enough value of $N$. These analytical results have
been numerically confirmed for uniform node sets in which nodes
are uniformly distributed in a two-dimensional space and for
fractal node sets in which nodes are placed in a fractal manner
\cite{Yakubo11}.

\subsection{Urban indicator}

In order to clarify the scaling property of an urban indicator $Y(N)$
quantifying activities in a city, we must relate $Y(N)$  to human
relations in the city modeled by a geographical network described
above. Although actual urban performances are sometimes produced by a
cooperation between many people in a group or an organization,
we consider here that the total urban performance stems from
one-to-one human relationships, namely from individual connected node
pairs in the network. Furthermore, we neglect nonlinear effects such as
interactions between individual node-pair activities creating additional
activities. These simplifications allows us to write the urban indicator
as
\begin{equation}
Y(N)=\frac{1}{2}\sum_{i,j}^{N} a_{ij}y_{ij}\ ,
\label{Y(N)_def}
\end{equation}
where $a_{ij}$ is the $(i,j)$ element of the adjacency matrix of the
network and $y_{ij}$ is the individual activity between nodes $i$ and
$j$.

As in the case of the AKS model \cite{Arbesman09} in which the individual
productivity is assumed to increase with the \textit{social} distance $d$,
it is natural to consider that the individual activity $y_{ij}$ depends on
the \textit{Euclidean} distance $l_{ij}$ between nodes $i$ and $j$.
Instead of the exponential $d$-dependence in the AKS model, we assume
a power-law dependence of $y_{ij}$ on $l_{ij}$, i.e.,
\begin{equation}
y_{ij}\propto l_{ij}^{\eta}\ ,
\label{yij_def}
\end{equation}
where the exponent $\eta$ can take either positive or negative values.
If $\eta$ is positive, longer-distance connections give higher individual
activities. In this case, we can regard $y_{ij}$ as an individual creative
productivity, because distant individuals have usually different
experiences and their values, and the fusion of heterogeneous ideas often
leads to greater creativity compared to combinations of homogeneous ideas.
This interpretation is consistent with the geographical network model
presented in the previous subsection. In the network model, a long-distance
connection is established only when two nodes have large attractiveness,
namely, they are highly capable. Outputs by collaboration between
such talented individuals must be innovative.

On the other hand, if $\eta$ is negative and $y_{ij}$ decreases with
$l_{ij}$, short-distance connections contribute more significantly to the
total urban indicator $Y(N)$ than long-distance ones. In this case, the
following consideration suggests that $Y(N)$ represents an infrastructure
volume. The degree of infrastructure development depends on how strong
the demand for the infrastructure is. Since infrastructure facility,
such as electrical power cables, railway stations, or green open urban spaces,
provides services for inhabitants near the facility, the social need
for the infrastructure arises from local consensus among neighboring
residents in areas having no access to the infrastructure. Thus, the
consensus between residents close to each other must be stronger than
that between distant ones. If we regard $y_{ij}$ given by Eq.~(\ref{yij_def})
with negative $\eta$ as the strength of the consensus between nodes
$i$ and $j$, $Y(N)$ provided by Eq.~(\ref{Y(N)_def}) quantifies the
whole social need in the city. Considering that infrastructure facilities
are realized in proportion to the social need, $Y(N)$ is proportional
to the infrastructure volume.

\section{URBAN SCALING}

In this work, we concentrate on the urban indicator averaged over
all possible cities with the same population size $N$ but different
spatial arrangements of people. Then, we treat the quantity,
\begin{equation}
Y(N)=\frac{1}{2}\left\langle \sum_{i,j}^{N} a_{ij}y_{ij}\right\rangle\ ,
\label{Y(N)_def2}
\end{equation}
where $\langle \cdots\rangle$ denotes the average over network
configurations with the same parameters $D$, $\alpha$, $m$, and $\eta$.
Using the node connection probability $g(l)$ by an edge of the length $l$,
the average urban indicator is presented by
\begin{equation}
Y(N)\propto N\int_{0}^{L} g(l)y(l)n(l)\,dl\ ,
\label{Y(N)_cal1}
\end{equation}
where $n(l)dl$ is the number of nodes within the range of $[l,l+dl]$
from a given node and $y(l)$ is the individual activity between nodes
separated each other by the distance $l$. In this section, we examine
the scaling behavior of $Y(N)$ by evaluating Eq.~(\ref{Y(N)_cal1}).

Substituting the relations $y(l)\propto l^{\eta}$ from Eq.~(\ref{yij_def}),
$n(l)\propto l^{D-1}$, and Eq.~(\ref{gl1}) into Eq.~(\ref{Y(N)_cal1}),
we have
\begin{eqnarray}
\frac{Y(N)}{N} &\propto:& \int_{0}^{\xi} l^{\eta} l^{D-1}\,dl+
\int_{\xi}^{L} \left(\frac{l}{\xi}\right)^{-d_{\text{c}}}l^{\eta} l^{D-1}\,dl \nonumber \\[5pt]
&\propto:& \xi^{D+\eta} +L^{D+\eta}\left(\frac{L}{\xi}\right)^{-d_{\text{c}}} ,
\label{Y(N)_cal2}
\end{eqnarray}
where the symbol $\propto:$ has been defined below Eq.~(\ref{kav2}).
Here we assumed
\begin{equation}
\eta>-D\ ,
\label{condition}
\end{equation}
for the convergence of the integral at $l=0$. This condition is,
however, not important because of the existence of the minimum node-pair distance
in actual spatial arrangements of people. Since the linear size $L$ is
related to $N$ through Eq.~(\ref{node_num_frac}), $Y(N)$ is written as
\begin{equation}
\frac{Y(N)}{N}\propto: \Theta^{-(D+\eta)/m}+\Theta^{-d_{\text{c}}/m}N^{1-(d_{\text{c}}-\eta)/D} ,
\label{Y(N)_cal3}
\end{equation}
where the characteristic length $\xi$ in Eq.~(\ref{Y(N)_cal2})
was replaced with the threshold $\Theta$ by using
Eq.~(\ref{xi}). Equation (\ref{Y(N)_cal3}) tells us how the
urban indicator scales with the population size $N$ under a
fixed value of the threshold $\Theta$.

We should note that as predicted by Eq.~(\ref{kav4}) the
average degree $\langle k\rangle$ of the network changes as $N$
increases under a fixed $\Theta$. In actual cities, however,
the average number of acquaintances corresponding to $\langle k\rangle$
is almost independent of $N$. Therefore, we must reveal the scaling
behavior of $Y(N)$ under a fixed value of $\langle k\rangle$
instead of a fixed $\Theta$. In order to express $Y(N)$ as a
function of $N$ and $\langle k\rangle$, we rewrite Eq.~(\ref{kav4})
as
\begin{subnumcases}
{\Theta \propto \label{Theta_kav}}
\label{Theta_kava}
\langle k\rangle^{-m/D} & $\text{if } D\le d_{\text{c}}$, \\[5pt]
\label{Theta_kavb}
\langle k\rangle^{-m/d_{\text{c}}}N^{m(D-d_{\text{c}})/Dd_{\text{c}}}
& $\text{if } D>d_{\text{c}}$.
\end{subnumcases}
In the case of $D\le d_{\text{c}}$, substitution of
Eq.~(\ref{Theta_kava}) into Eq.~(\ref{Y(N)_cal3}) yields
\begin{equation}
Y(N)\propto: \langle k\rangle^{1+\eta/D}N +\langle k\rangle^{d_{\text{c}}/D} N^{2+(\eta-d_{\text{c}})/D}.
\label{Y(N)_cal4}
\end{equation}
This relation is valid for a large enough population size,
because Eq.~(\ref{Theta_kav}) derived from Eq.~(\ref{kav4})
holds for a large $N$. In this case, one of two terms in
Eq.~(\ref{Y(N)_cal4}) dominates $Y(N)$ depending on the value
of the exponent of $N$. If $2+(\eta-d_{\text{c}})/D\le 1$,
namely $D\le d_{\text{c}}-\eta$, the first term grows with $N$
faster than the second term, and we have linear scaling of
$Y(N)$, i.e.,
\begin{equation}
Y(N)\propto N , \quad \text{if $D\le d_{\text{c}}$ and $D\le d_{\text{c}}-\eta$}.
\label{scaling_1}
\end{equation}
For $D> d_{\text{c}}-\eta$, however, the second term of Eq.~(\ref{Y(N)_cal4})
dominates $Y(N)$. Thus, $Y(N)$ scales as
\begin{equation}
Y(N)\propto N^{2+(\eta-d_{\text{c}})/D} , \quad \text{if $d_{\text{c}}-\eta<D\le d_{\text{c}}$}.
\label{scaling_2}
\end{equation}
On the other hand, for $D> d_{\text{c}}$, substitution of Eq.~(\ref{Theta_kavb})
into Eq.~(\ref{Y(N)_cal3}) leads to
\begin{eqnarray}
Y(N)\propto: \ &\langle k\rangle&^{(D+\eta)/d_{\text{c}}}N^{[d_{\text{c}}(2D+\eta)-D(D+\eta)]/Dd_{\text{c}}} \nonumber \\
&+&\langle k\rangle N^{1+\eta/D} .
\label{Y(N)_cal5}
\end{eqnarray}
Similarly to the case of Eq.~(\ref{Y(N)_cal4}), the comparison between
the exponents $[d_{\text{c}}(2D+\eta)-D(D+\eta)]/Dd_{\text{c}}$ and
$1+\eta/D$ gives
\begin{equation}
Y(N)\propto \ N^{[d_{\text{c}}(2D+\eta)-D(D+\eta)]/Dd_{\text{c}}} ,\quad \text{if $d_{\text{c}}<D\le d_{\text{c}}-\eta$},
\label{scaling_3}
\end{equation}
and
\begin{equation}
Y(N)\propto N^{1+\eta/D} ,\quad \text{if $D> d_{\text{c}}$ and $D> d_{\text{c}}-\eta$}.
\label{scaling_4}
\end{equation}
These relations provide nonlinear scaling of the urban indicator $Y(N)$.

Summarizing the above results, the scaling exponent $\beta$ in
Eq.~(\ref{urban_scaling}) is given by
\begin{widetext}
\begin{subnumcases}
{\beta = \label{exponent}}
\label{exponent_a}
1 & $\text{if } D\le d_{\text{c}} \text{ and } D\le d_{\text{c}}-\eta$ \\[3pt]
\label{exponent_b}
2+\frac{\eta-d_{\text{c}}}{D} & $\text{if } d_{\text{c}}-\eta<D\le d_{\text{c}}$ \\[3pt]
\label{exponent_c}
2+\frac{\eta}{D}-\frac{D+\eta}{d_{\text{c}}} & $\text{if } d_{\text{c}}<D\le d_{\text{c}}-\eta$ \\[3pt]
\label{exponent_d}
1+\frac{\eta}{D} & $\text{if } D>d_{\text{c}} \text{ and } D>d_{\text{c}}-\eta$\ .
\end{subnumcases}
\end{widetext}
The exponent $\beta$ can take any positive value by controlling
the four parameters $D$, $\alpha$, $m$, and $\eta$. This implies that the
urban indicator in our model scales superlinearly ($\beta>1$), linearly
($\beta=1$), or sublinearly ($\beta<1$) with the population size $N$.
Let us consider the value of $\beta$ by examining each expression of
Eq.~(\ref{exponent}). The exponent $\beta$ presented by Eq.~(\ref{exponent_a})
obviously leads to linear scaling of $Y(N)$. In this case, the exponent
$\eta$ can be positive or negative. If $\eta\ge 0$, the condition
for Eq.~(\ref{exponent_a}) is read as $D\le d_{\text{c}}-\eta$, namely,
$D+\eta\le m(\alpha-1)$, while it becomes $D\le d_{\text{c}}$ [i.e.,
$D\le m(\alpha-1)$] for $\eta<0$. Next, $\beta$ by Eq.~(\ref{exponent_b})
is always larger than $1$, because $(\eta-d_{\text{c}})/D$ is larger
than $-1$ from the condition $d_{\text{c}}-\eta<D$. We should note
that the condition for Eq.~(\ref{exponent_b}) requires $\eta> 0$.
On the contrary, Eq.~(\ref{exponent_c}) is the case
only when $\eta<0$. Taking into account Eq.~(\ref{condition}),
$\eta$ in Eq.~(\ref{exponent_c}) must satisfy $-D<\eta<0$ actually. Since
$(D+\eta)/d_{\text{c}}\le 1$ for Eq.~(\ref{exponent_c}), we have
$\beta\ge 1+\eta/D$. In addition, the condition $\eta>-D$ gives
$\beta>0$. Furthermore, $\beta$ given by Eq.~(\ref{exponent_c})
is expressed as $\beta=1+(D+\eta)(1/D-1/d_{\text{c}})$. Since
$D+\eta>0$ because of $\eta>-D$ and $(1/D-1/d_{\text{c}})<0$
because of $d_{\text{c}}<D$, the value of $\beta$ is less than $1$.
Therefore, the exponent $\beta$ presented by Eq.~(\ref{exponent_c})
can take a value in the interval $0<\beta<1$. Finally, for
Eq.~(\ref{exponent_d}), $\eta$ can be positive or negative. If
$\eta\ge 0$, obviously $\beta\ge 1$, whereas $0<\beta<1$ for
$-D<\eta<0$.

\begin{figure}[ttt!]
\begin{center}
\includegraphics[width=0.35\textwidth]{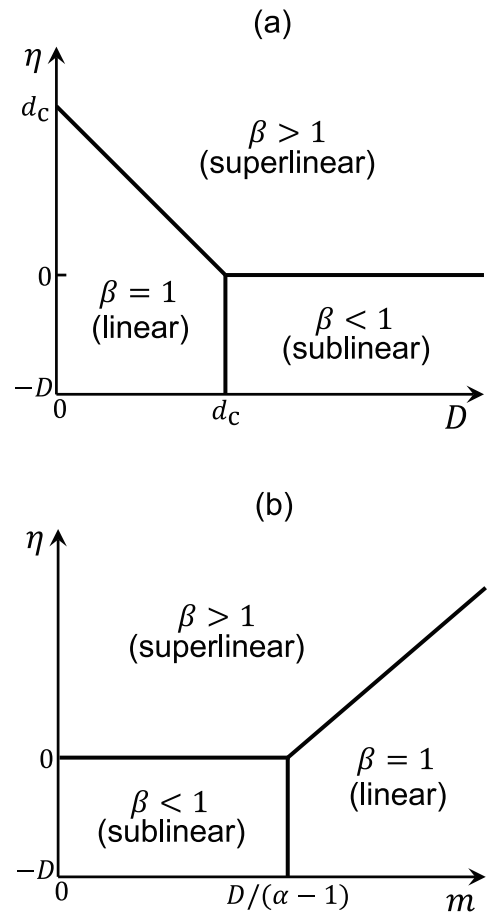}
\caption{Phase diagrams of our model (a) in the $D$-$\eta$ space with
fixed values of $m$ and $\alpha$ and (b) in the $m$-$\eta$ space with
fixed values of $D$ and $\alpha$. On the phase boundaries represented
by thick lines, $\beta$ is equal to $1$ (linear scaling).}
\label{fig:1}
\end{center}
\end{figure}
We can draw the phase diagram of our model from the above
results. Figure \ref{fig:1}(a) shows the regions of three
distinct scaling behaviors in the parameter space of $\eta$ and
$D$ under fixed values of $m$ and $\alpha$, and
Fig.~\ref{fig:1}(b) demonstrates those in the parameter space
of $\eta$ and $m$ under fixed values of $D$ and $\alpha$. The
phase boundaries in Fig.~\ref{fig:1}(b) are translated from
Fig.~\ref{fig:1}(a) by using Eq.~(\ref{dc}). These results
clearly show that superlinear scaling appears if $\eta$ is
positive and sublinear scaling if $\eta$ is negative. Since the
urban indicators $Y(N)$ constructed by positive and negative
$\eta$ correspond to a creative productivity and
infrastructure, respectively, these analytical results are
consistent with urban scaling observed in the real world
\cite{Bettencourt07}. Note that we have linear scaling
($\beta=1$) on the phase boundaries. Thus, the condition
$\eta=0$ always gives linear scaling regardless of the values
of other parameters. This is reasonable because $Y(N)$ for
$\eta=0$ is nothing but the number of edges $M$ in the network
and $M$ is proportional to $N$ when $\langle k\rangle$ is
independent of $N$. The urban indicator that scales linearly
corresponds to individual human needs such as the total number
of houses.

It is found from Fig.~\ref{fig:1}(b) that $Y(N)$ always obeys
linear scaling for large enough $m$, i.e., $\beta=1$ when the
geographical constraint in the network formation is very
strong. Since the network formed by a large $m$ value has a
lattice-like structure as mentioned below Eq.~(\ref{dc}),
lengths of edges in the network are almost constant. This is
also confirmed by the fact that the edge-length distribution
$R(l)$ given by Eq.~(\ref{kl1}) becomes narrower as $m$
increases. If edge lengths are constant, individual node-pair
activities given by $y_{ij}\propto l_{ij}^{\eta}$ are also
constant. Denoting this constant by $y_{0}$,
Eq.~(\ref{Y(N)_def2}) provides $Y(N)=N\langle k\rangle y_{0}/2$,
which leads to linear scaling.

\section{SIMULATION RESULTS}

\begin{figure}[ttt!]
\begin{center}
\includegraphics[width=0.4\textwidth]{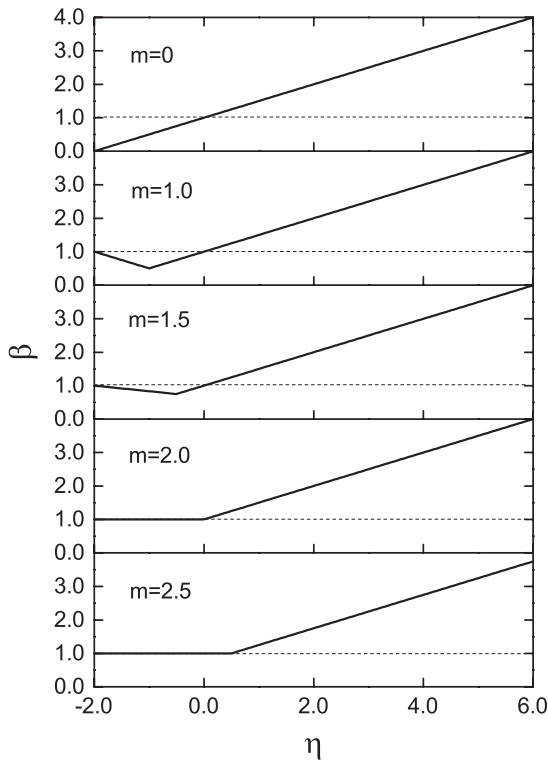}
\caption{Urban scaling exponent $\beta$ as a function of $\eta$ for
several values of $m$. The exponents $\alpha$ and the fractal dimension
$D$ are fixed at $\alpha=2.0$ and $D=2.0$. Horizontal dashed lines
at $\beta=1$ are guides to the eye, which separate the superlinear
scaling region from the sublinear one.}
\label{fig:2}
\end{center}
\end{figure}
\begin{figure*}[ttt!]
\begin{center}
\includegraphics[width=0.85\textwidth]{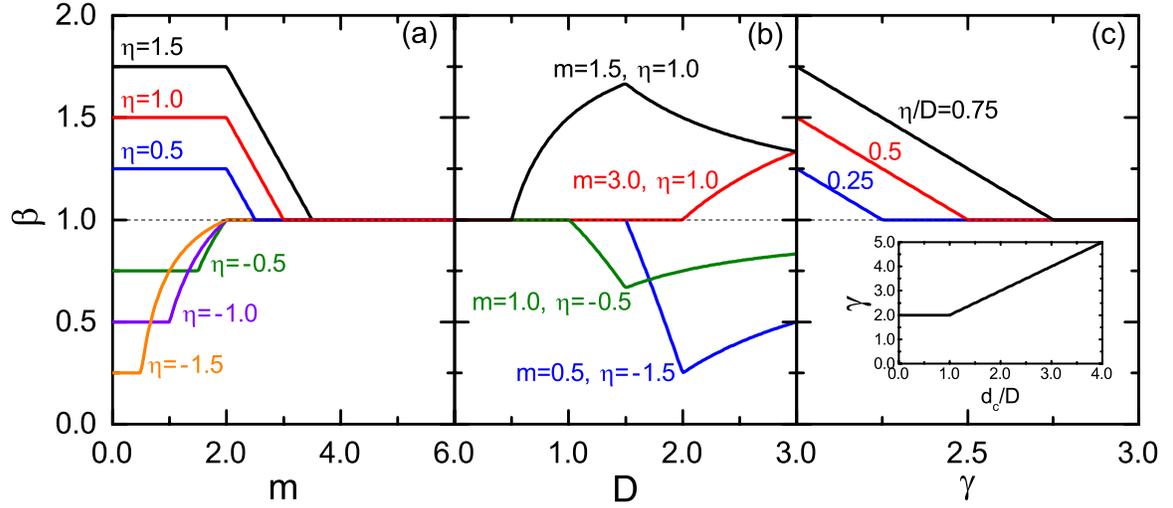}
\caption{(Color online) Profiles of the urban scaling exponent
$\beta$ as a function of $m$, $D$, and $\gamma$. (a) $\beta$
versus $m$ for various values of $\eta$. The exponents $\alpha$
and $D$ are fixed as $\alpha=2.0$ and $D=2.0$. (b) $\beta$
versus $D$ for various combinations of $m$ and $\eta$. The
exponent $\alpha$ is fixed at $\alpha=2.0$. (c) $\beta$ versus
$\gamma$ for various positive values of $\eta/D$. Dashed line
at $\beta=1$ in each panel separates the superlinear scaling
region from the sublinear one. The inset of (c) shows the
$d_{\text{c}}/D$ dependence of the exponent $\gamma$ given by
Eq.~(\ref{gamma}).}
\label{fig:3}
\end{center}
\end{figure*}
The urban scaling exponent $\beta$ predicted by our model depends on the
parameters $D$, $\alpha$, $m$, and $\eta$. Typical profiles of $\beta$
given by Eq.~(\ref{exponent}) are presented in Figs.~\ref{fig:2} and \ref{fig:3}.
Figure \ref{fig:2} shows the $\eta$ dependence of $\beta$ for various values
of $m$ under $D=2.0$ and $\alpha=2.0$. It is verified that superlinear scaling
of $Y(N)$ requires $\eta>0$ and sublinear scaling is allowed only for $\eta<0$.
For any combination of $m$, $\alpha$, and $D$, the exponent $\beta$ linearly
increases with $\eta$ if $\eta$ is large enough. The $m$ dependence of $\beta$
is depicted in Fig.~\ref{fig:3}(a) for various values of $\eta$. This figure
clearly demonstrates that the urban indicator scales linearly with the population
size if $m$ is large enough, as pointed out in the previous section. The fact
of $\beta\ne 1$ at $m=0$ shows that the geographical constraint
\textit{in the network formation} is not necessary for nonlinear urban scaling,
which does not mean, however, that networks are not required to be embedded
in the Euclidean space for obtaining nonlinear scaling of $Y(N)$.
The exponent $\beta$ changes with the fractal dimension $D$ as shown in
Fig.~\ref{fig:3}(b). In contrast to the $m$ dependence, $\beta$ depends
non-monotonically on $D$. Although only results for $\eta<d_{\text{c}}$
are shown here, $\beta$ for $\eta>d_{\text{c}}$ monotonically decreases
with $D$ and diverges at $D=0$. Despite the lack of a physical meaning
of the divergent $\beta$ in the limit of $D=0$, a large value of $\beta$
at small $D$ is reasonable because the system must have very long
edges to keep $\langle k\rangle$ constant and $Y(N)$ increases rapidly
with $N$.

Since the exponent $\gamma$ characterizing the scale-free property of
the network depends on $d_{\text{c}}$ and $D$ as presented by
Eq.~(\ref{gamma}), it seems interesting to elucidate how the urban scaling
exponent $\beta$ varies with $\gamma$. The model parameter $d_{\text{c}}$
giving $\gamma=2$ for a fixed $D$ is, however, not uniquely determined
if $D\ge d_{\text{c}}$ [see the inset of Fig.~\ref{fig:3}(c)]. Thus,
$\beta$ for sublinear scaling that requires $D\ge d_{\text{c}}$ cannot
be related to $\gamma$. On the other hand, there is a one-to-one
correspondence between $\gamma$ and $d_{\text{c}}$ for a fixed $D$ if
$D< d_{\text{c}}$ that leads to superlinear or linear scaling. In this
case,  from Eqs.~(\ref{exponent_a}) and (\ref{exponent_b}), the exponent
$\beta$ is expressed as
\begin{equation}
\beta =
\begin{cases}
\displaystyle
3+\frac{\eta}{D}-\gamma & \text{if } 2<\gamma<2+\displaystyle\frac{\eta}{D} \ ,\\[5pt]
\displaystyle
1  & \text{if } \gamma\ge 2+\displaystyle\frac{\eta}{D}\ ,
\end{cases}
\label{exp_gamma}
\end{equation}
where $\eta$ must be positive. The $\gamma$ dependence of $\beta$ for
$\eta>0$ and $\gamma>2$ is illustrated in Fig.~\ref{fig:3}(c). From this
argument, we can conclude that sublinear scaling is realized in a network
with $\gamma=2$ and superlinear scaling appears for $2<\gamma<2+\eta/D$
in our model.

\begin{figure}[bbb!]
\begin{center}
\includegraphics[width=0.45\textwidth]{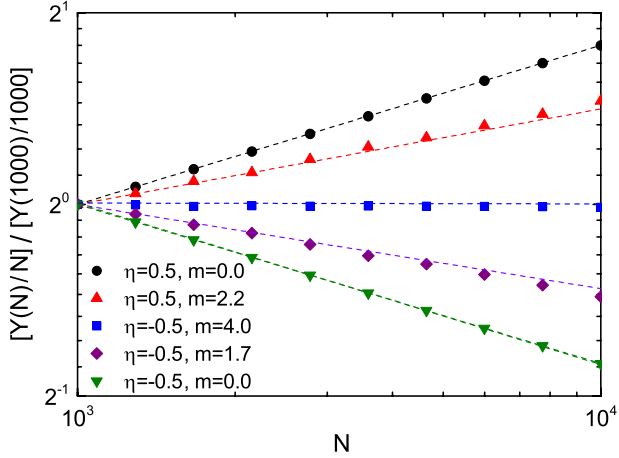}
\caption{(Color online) Numerically calculated urban indicators
as a function of the population size (number of nodes) $N$.
Nodes in geographical networks are scattered uniformly at
random in two-dimensional square spaces ($D=2$) with a fixed
node density. The exponent $\alpha$ and the parameter
$x_{\text{min}}$ characterizing the attractiveness distribution
given by Eq.~(\ref{sx}) are chosen as $\alpha=2.0$ and
$x_{\text{min}}=1.0$. The longitudinal axis indicates $Y(N)/N$
rescaled by its value at $N=1,000$. Each symbol represents the
result averaged over $1,000$ realizations. Standard errors are
smaller than the size of symbols. Circles, triangles, squares,
diamonds, and inverted triangles are the results for
($\eta=0.5$, $m=0$), ($\eta=0.5$, $m=2.2$), ($\eta=-0.5$,
$m=4.0$), ($\eta=-0.5$, $m=1.7$), and ($\eta=-0.5$, $m=0$),
respectively. Dashed lines through symbols from the top to
the bottom give the theoretically predicted slopes of
$\beta-1=0.25$, $0.155$, $0.0$, $-0.132$, and $-0.25$,
respectively.}
\label{fig:4}
\end{center}
\end{figure}
Let us confirm the above analytical results by numerical
simulations. For simplicity, we treat the case of $D=2$,
namely, $N$ nodes are uniformly distributed at random in a
two-dimensional square space. The linear size $L$ of the
square space is adjusted to keep the node density constant with
a change in $N$. The attractiveness $x_{i}$ is assigned to each
node according to the distribution function Eq.~(\ref{sx}) with
$\alpha=2.0$ and $x_{\text{min}}=1.0$. The Euclidean distance
$l_{ij}$ between nodes $i$ and $j$ is measured under periodic
boundary conditions, and the threshold value $\Theta$ in
Eq.~(\ref{condition1}) is chosen so that the average degree
becomes $\langle k\rangle=10.0$. Networks formed by these
conditions possess the scale-free property characterized by
$\gamma=2$ for $m\le 2$ and $\gamma=1+m/2$ for $m>2$. The urban
indicator $Y(N)$ is calculated directly from the definition
Eq.~(\ref{Y(N)_def2}). Figure \ref{fig:4} shows the $N$
dependence of $Y(N)$ for various combinations of $\eta$ and $m$
in a double logarithmic scale. The longitudinal axis represents
$Y(N)/N$ rescaled by its value at the minimum $N$ ($=1,000$) to
improve the legibility of the results. Thus, an increasing,
decreasing, or constant straight line indicates superlinear,
sublinear, or linear scaling of $Y(N)$, respectively. Our
numerical results clearly show that $Y(N)$ obeys a power law
with respect to $N$ and the slopes representing $\beta-1$ agree
with the theoretical predictions indicated by dashed lines.
Triangles ($\eta=0.5$ and $m=2.2$) and diamonds ($\eta=-0.5$
and $m=1.7$) in Fig.~\ref{fig:4} slightly deviate from the
corresponding theoretical lines. These deviations are caused by
the finite-size effect as discussed below.

\begin{figure}[ttt!]
\begin{center}
\includegraphics[width=0.45\textwidth]{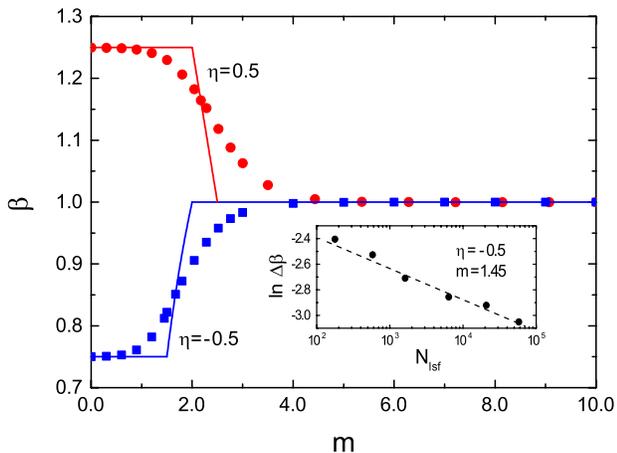}
\caption{(Color online) Numerically calculated $m$ dependence
of the exponent $\beta$. Circles and squares represent the
results for $\eta=0.5$ and $-0.5$, respectively. All the
conditions other than $m$ and $\eta$ are the same with those for
Fig.~\ref{fig:4}. Solid lines give the theoretical predictions
by Eq.~(\ref{exponent}) for $\eta=0.5$ and $-0.5$ ($\alpha=2.0$
and $D=2.0$ for both lines). The inset shows the deviation
$\Delta \beta$ of $\beta$ calculated numerically for $\eta=-0.5$
and $m=1.45$ from its theoretical value as a function of
$N_{\text{lsf}}$ around which the least squares fit is performed
within a narrow window of $N$. Dashed line in the inset is a
guide to the eye.}
\label{fig:5}
\end{center}
\end{figure}

Next, we numerically calculated values of $\beta$ as a function
of $m$ and compare the obtained results with the theoretical
predictions. The exponent $\beta$ is estimated by the least
squares fit for numerical data of $Y(N)$ within the range of
$10^{3}\le N \le 10^{4}$. Results for $\eta=0.5$ and
$\eta=-0.5$ are presented by filled circles and squares in
Fig.~\ref{fig:5}, respectively. Parameters other than $\eta$
and $m$ and the computational conditions, such as the boundary
conditions and the number of realizations for the sample
average, are the same as those for Fig.~\ref{fig:4}. Standard
errors over samples are less than the symbol size. Solid lines
in Fig.~\ref{fig:5} represent the theoretical predictions given
by Eq.~(\ref{exponent}) for $\eta=0.5$ and $-0.5$. Numerical
results roughly coincide with the theoretical curves.
Especially, data for $m>4$ and $m<1$ agree quite well with the
theoretical curves. However, simulation results near
$m=D/(\alpha-1)$ and $(D+\eta)/(\alpha-1)$ that give the
turnoff points of $\beta(m)$ (i.e., $m=2.0$ and $2.5$ for
$\eta=0.5$ and $m=2.0$ and $1.5$ for $\eta=-0.5$) deviate from
the theoretical values. This is due to the finite-size effect.
In the analytical calculation of the exponent $\beta$, we
assume a large enough number of nodes to determine the dominant
terms of Eqs.~(\ref{kav3}), (\ref{Y(N)_cal4}), and
(\ref{Y(N)_cal5}). If two exponents of $N$ in each of these
equations becomes close to each other (i.e., approaching to the
turnoff point), both terms almost equally contribute to $Y(N)$
[or to $\langle k\rangle$ for Eq.~(\ref{kav3})], and $Y(N)$ for
numerically accessible $N$ does not obey a power law any more.
In order to demonstrate that the deviation of numerically
calculated $\beta$ near the turnoff point is caused by the
finite-size effect, we show the network-size dependence of the
deviation $\Delta \beta$ of numerical data from the theoretical
one in the inset of Fig.~\ref{fig:5}. For obtaining this inset,
we calculated numerically $Y(N)$ for $\eta=-0.5$ and $m=1.45$
within the range of $10^{3}\le N \le 10^{5}$ and estimated
$\beta$ by the least squares fit for these data in relatively
narrow windows of $N$ around $N_{\text{lsf}}$. The result in
the inset displays that the deviation $\Delta \beta$ decreases
with increasing $N_{\text{lsf}}$, which suggests $\Delta
\beta=0$ in the thermodynamic limit ($N\to \infty$).

\section{CONCLUSION}

The origin of superlinear and sublinear scaling observed in
urban indicators has been analytically argued by modeling the
interrelationship of people in a city by a geographical
scale-free network. In this network model, nodes close to each
other are more likely to be connected than long distant nodes.
We assumed that the urban indicator $Y$ of a city is given by
the sum of individual node-pair activities $\{y_{ij}\}$
produced by personal, one-to-one human relationships in the city
and $y_{ij}$ is proportional to $l_{ij}^{\eta}$, where $l_{ij}$
is the Euclidean distance between directly connected nodes $i$
and $j$. For a positive or negative exponent $\eta$, the urban
indicator represents a creative productivity or a degree of
infrastructure development, respectively. We showed that the
urban indicator obeys a power law $Y(N)\propto N^{\beta}$ for a
large enough population size $N$. The exponent $\beta$ is
larger than or equal to one if $\eta>0$, while $0< \beta\le 1$
for $\eta<0$, which implies that $Y(N)$ corresponding to a
creative productivity scales superlinearly or linearly with
respect to the population size $N$ and it scales sublinearly or
linearly if $Y(N)$ is a quantity related to infrastructure.
This result coincides with the scaling behavior of real-world
urban indicators. It has been also found that $Y(N)$ is
proportional to $N$ if networks are formed under a strong
geographical constraint. These results have been confirmed
by numerical simulations.

In our argument, nodes are assumed to be placed on a $D$
dimensional Euclidean space and the geographical distance plays
a crucial role to understand urban scaling. To interpret $Y(N)$
under a negative $\eta$ as a degree of infrastructure
development, the nodes must be arranged in a physical
(geographic) space. This condition, however, can be relaxed for
superlinear scaling. We can derive the same result for
superlinear scaling of $Y(N)$ even in the case that nodes are
placed on a more general metric space in which
Eqs.~(\ref{condition1}) and (\ref{yij_def}) with the abstract
distance $l_{ij}$ are a reasonable condition for the network
formation and a plausible relation for the individual activity,
respectively. For example, in a sociometric space, where social
distances between nodes are defined, we can consider that nodes
socially close to each other are more likely to be connected
and a socially more distant node pair yields a higher
productivity. Therefore, the scaling exponent $\beta$ is also
presented by Eq.~(\ref{exponent}) for $\eta>0$, if
Eqs.~(\ref{condition1}) and (\ref{yij_def}) with the social
distance $l_{ij}$ do actually hold. We should note that in such
a case $D$ must be the (fractal) dimension of the sociometric
space.

We concentrated, in this work, on the average scaling behavior
of the urban indicator. However, the actual urban indicators of individual
cities deviate from the average values of $Y(N)$
expected from their population sizes. Statistical properties of
the fluctuations of $Y(N)$ have been extensively studied by
recent works \cite{Bettencourt10,Gomez-Lievano12,Lobo13,Alves13}.
Within the framework of the present model, we can also consider
such fluctuations by evaluating $Y(N)$ defined by
Eq.~(\ref{Y(N)_def}) instead of its average given by
Eq.~(\ref{Y(N)_def2}). The fluctuations in the urban indicators
of cities with the same population size $N$ are caused, in our
model, by different network configurations due to different
spatial arrangements of nodes and different assignments of the
attractiveness. In addition to this structural network effect,
the deviation of $Y(N)$ from the average value could arise from
the fluctuations in the model parameters. There are four
parameters in our model, i.e., $m$ characterizing the strength
of the geographical constraint in the network formation,
$\alpha$ describing how widely distributed the attractiveness
is, the fractal dimension $D$ of the population density, and
$\eta$ specifying the Euclidean-distance dependence of the
individual activity. Although this work assumes that these four
parameters remain constant over a set of cities, violation of
this assumption will also lead to fluctuations in the urban
indicator. By comparing statistical properties of the predicted
fluctuations of $Y(N)$ to those observed in actual urban
indicators we would be able to assess how well our model
describes urban scaling phenomena.

\begin{acknowledgments}
This work was supported by a Grant-in-Aid for Scientific
Research (No.~22560058) from Japan Society for the Promotion of
Science, by the project SEETechnology  ``Co-operation of SEE
science parks for the promotion of transnational market uptake
of R\&D results and technologies by SMEs'' co-funded by South
East Europe Transnational Cooperation Programme, and by the
operation entitled ``Centre for Open Innovation and Research of
the University of Maribor''. The last operation is co-funded by
the European Regional Development Fund and conducted within the
framework of the Operational Programme for Strengthening
Regional Development Potentials for the period 2007--2013,
development priority 1: ``Competitiveness of companies and
research excellence'', priority axis 1.1: ``Encouraging
competitive potential of enterprises and research excellence''.

Numerical calculations in this work were performed in part on
the facilities of the Supercomputer Center, Institute for Solid
State Physics, University of Tokyo.
\end{acknowledgments}

\end{document}